# Fast Algorithms for Mining Interesting Frequent Itemsets without Minimum Support

Shariq Bashir, Zahoor Jan, A. Rauf Baig

FAST-National University of Computer and Emerging Science

A. K. Brohi Road, H-11/4, Islamabad, Pakistan

{shariq.bashir, i040930, rauf.baig}@nu.edu.pk

## ABSTRACT

Real world datasets are sparse, dirty and contain hundreds of items. In such situations, discovering interesting rules (results) using traditional frequent itemset mining approach by specifying a user defined input support threshold is not appropriate. Since without any domain knowledge, setting support threshold small or large can output nothing or a large number of redundant uninteresting results. Recently a novel approach of mining only N-most/Top-K interesting frequent itemsets has been proposed, which discovers the top N interesting results without specifying any user defined support threshold. However, mining interesting frequent itemsets without minimum support threshold are more costly in terms of itemset search space exploration and processing cost. Thereby, the efficiency of their mining highly depends upon three main factors (1) Database representation approach used for itemset frequency counting, (2) Projection of relevant transactions to lower level nodes of search space and (3) Algorithm implementation technique. Therefore, to improve the efficiency of mining process, in this paper we present two novel algorithms called (N-MostMiner and Top-K-Miner) using the bit-vector representation approach which is very efficient in terms of itemset frequency counting and transactions projection. In addition to this, several efficient implementation techniques of N-MostMiner and Top-K-Miner are also present which we experienced in our implementation. Our experimental results on benchmark datasets suggest that the N-MostMiner and Top-K-Miner are very efficient in terms of processing time as compared to current best algorithms BOMO and TFP.

## 1. INTRODUCTION

Since the introduction of association rules mining by Agrawal et al. [1], it has now become one of the main pillars of data mining and knowledge discovery tasks and has been successfully applied in many interesting association rules mining problems such as sequential pattern mining [2], emerging pattern mining [8], classification [13], maximal and closed itemset mining [3, 5, 14]. Using the support-confidence framework presented in [1], the problem of mining the complete association rules from transactional dataset is divided into two parts – (a) finding complete frequent itemsets with support (an itemset's occurrence in the dataset) greater than minimum support threshold, (b) generating association rules from frequent itemsets with confidence greater than minimum confidence threshold. In practice, the first phase is the most time-consuming task, which requires the heaviest frequency counting operation for each candidate itemset.

Let *TDS* be our transactional dataset and *I* be a set of distinct items in the *TDS*. Each individual transaction *t* in *TDS* consists a subset of single items such as $t_I \subseteq I$. We call *X* an itemset, if it contains $X \subseteq I$. Let *min-sup* be our minimum support threshold, we call an itemset *X* frequent if its support *(support(X))* is greater than *min-sup*; otherwise infrequent. By following the Apriori property [1] an itemset *X* cannot be a frequent itemset, if one of its subset is infrequent. We denote the set of all frequent itemset by FI. If *X* is frequent and no superset of *X* is frequent, we say that *X* is a maximal frequent itemset, the set of all maximal frequent itemsets is denoted by MFI. If *X* is frequent and no superset of *X* is as frequent as *X*, we say that *X* is a closed frequent itemset; similarly the set of all closed frequent itemset is denoted by FCI. Thus the following condition is straight-forward holds: MFI $\subseteq$ FCI $\subseteq$ FI. The frequent itemsets mining algorithms take a transactional dataset *(TDS)* and *min-sup* as an input and output all those itemsets which appear in at least *min-sup* number of transactions in *TDS*. However, the real datasets are sparse, dirty and contain hundreds of items. In such situations, users face difficulties in setting this *min-sup* threshold to obtain their desired results. If *min-sup* is set too large, then there may be a small number of frequent itemsets, which does not give any desirable result. If the *min-sup* is set too small, then there may be a large number of redundant short uninteresting itemsets, which not only takes a large processing time for mining but also increases the complexity of filtering un-interesting itemsets. In both situations, the ultimate goal of mining interesting frequent itemsets is undermined. We refer the readers [7] for further reading about the problem of setting this user defined *min-sup* threshold without any previous domain knowledge about the dataset.

For handling such situations, Fu et al. in [9] presents a novel technique of mining only N-most interesting frequent itemsets without specifying any *min-sup* threshold. The problem of mining N-most interesting frequent itemsets of size $k$ at each level of $1 \leq k \leq k_{max}$, given $N$ and $k_{max}$ can be considered from the following definitions.

| TDS | The given transactional dataset |
|---|---|
| $k_{max}$ | Upper bound on the size of interesting itemsets to be found |
| $\xi$ | Current support threshold for all the itemsets |
| $\xi_k$ | Current support threshold for the k-itemsets |

**Definition 1:** *k-itemsets:* Is a set of items containing k items.

**Definition 2:** *N-most interesting k-itemsets:* Let *I* be the set of sorted itemsets by descending support. Let *S* be the support of the N-th *k*-itemset in the set *I*. The N-most interesting *k*-itemsets in the set *I* are those itemsets having *support* $\geq S$.

**Definition 3:** *The N-most interesting itemsets:* Is the union of the N-most interesting *k*-itemsets for each $1 \leq k \leq k_{max}$.

Han et al. in [18], proposed an another variation of mining Top-*K* frequent closed itemsets with length greater than a minimum user specified threshold *min_l*, where *K* is a user-desired number of frequent closed itemsets to be mined. Their work is different from [9] in this sense, that if a itemset *X* is found infrequent at any node *n*, then all the supersets of *X*, or the subtree of *n* can be safely pruned away, which decreases the overall processing time.

### 1.1. MOTIVATION BEHIND OUR WORK

As clear from the Definition 2 described above, the Apriori property presented by Agrawal et al. in [1] can not be applied in mining N-most interesting frequent itemset algorithm for pruning un-interesting itemsets. Since the superset of any uninteresting *k*-itemset may be the N-most interesting itemset of any level *h such that* $1 \leq h \leq k_{max}$. Therefore, a large area of itemset search space is explored as compared to traditional frequent itemset mining approach. In our different computational experiments on several sparse and dense benchmark datasets, we found that the efficiency of mining interesting frequent itemsets without minimum support threshold highly depends upon three main factors. (1) Dataset representation approach used for frequency counting [10]. (2) Projection of relevant transactions to lower level nodes of search space, and (3) Algorithm implementation technique. The projection of relevant transactions at any node *n*, are those transactions of dataset which contain the node *n*'s itemset as subset [11]. Therefore, to increase the efficiency of mining interesting frequent itemsets (N-Most or Top-K), in this paper we present novel efficient algorithms (N-MostMiner and Top-K-Miner) using Bit-vector dataset representation approach. The major advantage of using Bit-vector dataset representation approach in our algorithms is that, it optimizes the itemset frequency counting cost with a factor of 1/32, if we represent 32 rows per single vertical bit-vector region [17]. Before our work, Bit-vector dataset representation approach has been successfully applied in many complex association rules problems such as maximal frequent itemsets mining [6], sequential patterns mining [4] and fault-tolerant frequent itemsets mining [12]. In addition to dataset representation approach, this paper also presents a novel bit-vector projection technique which we named as projected-bit-regions (PBR). The main advantage of using PBR in N-MostMiner and Top-K-Miner is that, it consumes a very small processing cost and memory space for projection. In section 5 we also present some efficient implementation techniques of N-MostMiner and Top-K-Miner, which we experienced in our implementation. Our different experiments on benchmark datasets suggest that mining interesting frequent itemsets without minimum support threshold using our algorithms are fast and efficient than the currently best algorithms BOMO [7] and TFP [18].

## 2. RELATED WORK

The Apriori algorithm by Agrawal et al. [1] is considered as one of most well known algorithm in frequent itemset mining problem. Apriori uses a complete, bottom-up search, with a horizontal layout and prune infrequent itemsets using anti-monotone Apriori heuristic: if any length *k* pattern is not frequent in the database, its length *(k+1)* super pattern can never be frequent. Apriori is an iterative algorithm that counts itemsets of specific length in a given database pass. The three main steps of Apriori algorithm are as fallow:

1. Generating frequent candidate (k+1)-itemsets, by joining the frequent itemsets of previous pass k.
2. Deleting those subsets which are infrequent in the previous pass $k$ without considering the transactions in the dataset.
3. Scanning all transactions to obtain frequent candidate (k+1)-itemsets.

Although, Apriori presented by Agrawal et al. [1] is very effective method for enumerating frequent itemsets of sparse datasets on large support threshold, but the basic algorithm of Apriori encounters some difficulties and takes large processing time on low support threshold. We list here some main deficiencies of Apriori that make it an unattractive solution for mining frequent itemsets.

1. Apriori encounters difficulty in mining long pattern, especially for dense datasets. For example, to find a frequent itemsets of $X = \{1...200\}$ items. Apriori has to generate-and-test all $2^{200}$ candidates.
2. Apriori algorithm is considered to be unsuitable for handling frequency counting, which is considered to be most expensive task in frequent itemsets mining. Since Apriori is a level-wise candidate-generate-and-test algorithm, therefore it has to scan the dataset *200* times to find a frequent itemsets $X = X_1... X_{200}$.
3. Even though Apriori prunes the search space by removing all *k* itemsets, which are infrequent before generating candidate frequent *(k+1)*-itemsets, it still needs a dataset scan, to determine which candidate *(k+1)* itemsets are frequent and which are infrequent. Even for datasets which have *200* items, determining *k*-frequent itemsets by repeated scanning the dataset with pattern matching takes a large processing time.

Apriori and its variants enumerate all frequent itemsets by repeatedly scanning the dataset and checking the frequency of candidate frequent *k* itemsets by pattern matching. This whole process is costly especially if the dataset is dense and has long patterns, or a low minimum support threshold is given. To increase the efficiency of frequent itemset mining Han et al. [11] presented a novel method called pattern-growth. The major advantage of frequent itemsets mining using pattern-growth approach is that it removes the costly operation of repeatedly scanning the dataset in each iteration and generating and testing infrequent candidate itemsets. In simple words pattern growth removes the costly candidate-generate-and-test operation of Apriori type algorithms. The method requires only two dataset scans for counting all frequent itemsets. The first dataset scan collects the support of all frequent items. The second dataset scan builds a compact data structure called FP-tree. Each node of FP-tree corresponds to an item which was found frequent in first dataset scan. Next, all frequent itemsets are mined directly from this FP-tree without concerning the transactional dataset.

The pattern-growth approach mines all frequent itemsets on the basis of FP-tree. The main strategy of pattern-growth approach is that it traverses search space in depth first order, and on each node of search space it mines frequent itemsets on the basis of conditional patterns and creates child FP-tree (also called projected database).

Burdick et al. [6] presented a simple hardware efficient dataset representation scheme (vertical bit-vector) and a frequent itemset mining algorithm called MAFIA. To count the frequency of any itemset, the algorithm performs a simple bitwise-AND operation on two bit-vectors and resulting ones represents the frequency of that itemset. On 32-bit CPU the algorithms performs a bitwise-AND operation on 32-bits per operation. Mafia also used parent equivalence pruning (PEP) and differentiates superset pruning into two classes FHUT and HUTMFI. For a given node $X:aY$, the idea of PEP is that if $sup(X)=sup(Xa)$, i.e. every transaction containing $X$ also contains the item $a$, then the node can simply be replaced by $Xa:Y$. The FHUT uses leftmost tree to prune its sister, i.e., if the entire tree with root $Xa:Y$ is frequent, then we do not need to explore the sisters of the node $Xa:Y$. The HUTMFI uses to use the known MFI set to prune a node, i.e., if itemset of $XaY$ is subsumed by some itemset in the MFI set, the node $Xa:Y$ can be pruned. MAFIA also uses dynamic reordering to reduce the search space. The results show that PEP has the biggest effect of the above pruning methods (PEP, FHUT, and HUTMFI).

The idea of mining N-most interesting itemsets without specifying any input support threshold directly from the user was first proposed in Itemset-Loop algorithm [9]. The Itemset-Loop applies a variation of the Apriori-gen algorithm repeatedly, using different support thresholds and a modified candidate generation mechanism. Although the Itemset-Loop was first algorithm in N-most interesting itemsets mining category, it makes use of the Apriori candidate generation mechanism which relies on the property of subset closure: if a k-itemset is large then all of its subset will be also large. This property does not hold for mining N-most interesting itemsets. That is, if a k-itemset is among the N-most interesting k-itemsets, its subsets may not be among the N-most interesting itemsets.

BOMO algorithm presented in [7] is a frequent pattern-growth (FP-growth) based approach and known as the currently best algorithm in mining N-most interesting itemsets category. BOMO uses a compact frequent pattern-tree (FP-tree) to store compressed information about frequent itemsets. FP-growth is a depth first search based approach, and does not rely on candidate generation-and-test mechanism and achieves impressive results in frequent itemset mining problems.

Han et al. in [18], proposed an another algorithm, TFP, a pattern growth based approach, of mining Top-K interesting frequent closed itemset with length greater than *min_l* threshold, where *k* is the user-desired number of frequent closed itemsets to be mined. TFP starts with *min_support = 0*, and uses its efficient techniques for pruning FP-Tree dynamically both during and after the construction of the tree. At the time of initial FP-Tree construction, a closed node count array and descendent sum is maintained which raise minimum support before Top-K closed itemset mining. After initial FP-Tree construction, TFP uses the top down FP-Tree mining techniques to first mine the most promising parts of the tree in order to raise minimum support and prune the unpromising part of the tree during itemset mining. Moreover, TFP execution is further speed up by efficient itemset closure checking scheme and set of search space pruning methods.

| Transaction | Items |
|---|---|
| 01 | A B C |
| 02 | A B I |
| 03 | B |
| 04 | C I |
| 05 | A B D |
| 06 | A B C D |
| 07 | A |

(a)

| Transaction | A | B | C | D | I |
|---|---|---|---|---|---|
| 01 | 1 | 1 | 1 | 0 | 0 |
| 02 | 1 | 1 | 0 | 0 | 1 |
| 03 | 0 | 1 | 0 | 0 | 0 |
| 04 | 0 | 0 | 1 | 0 | 1 |
| 05 | 1 | 1 | 0 | 1 | 0 |
| 06 | 1 | 1 | 1 | 1 | 0 |
| 07 | 1 | 0 | 0 | 0 | 0 |

(b)

**Figure 1:** (a) A sample transactional dataset. (b) Bit-vector representation of sample dataset.

## 3. ALGORITHMIC DESCRIPTION

In this section, we describe our N-most interesting itemset mining algorithm (N-MostMiner) with its several techniques used for fast itemset frequency counting and projection. For *k*-itemset representation at any node, the N-MostMiner uses the vertical bit-vector representation approach [6]. In a vertical bit-vector representation, there is one bit for each transaction of dataset. If item *i* appears in transaction *j*, then the bit *j* of bit-vector *i* is set to one; otherwise the bit is set to zero. In Figure1 (a) a dataset is shown along with its vertical bit-vector representation in Figure1 (b). To count the frequency of an *k*-itemset e.g. *{AB}* we need to perform a bitwise-AND operation on bit-vector *{A}* and bit-vector *{B}*, and resulting ones in bit-vector *{AB}* denotes the frequency of *k*-itemset *{AB}*.

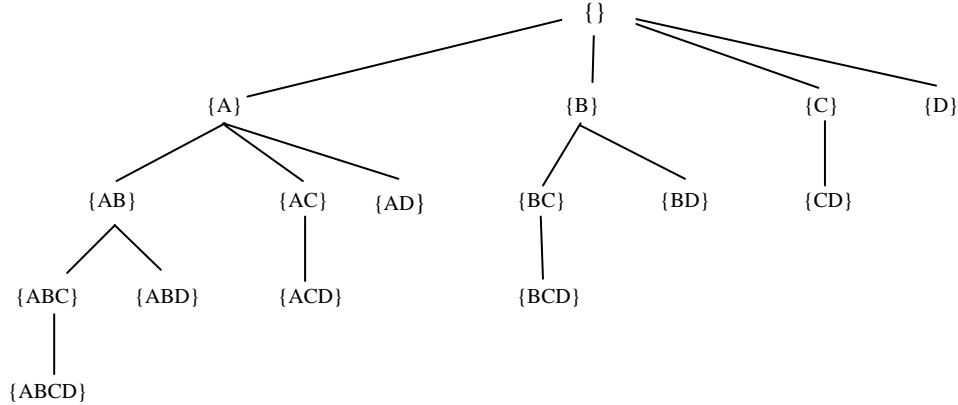

**Figure 2:** Lexicographic tree of four items ⟨A, B, C, D⟩.

### 3.1. ITEMSET GENERATION

Let < be some lexicographical order of items in Transactional Dataset (TDS) such that for every two items *a* and *b*, *a ≠ b: a < b or a > b*. The search space of mining N-most interesting itemset mining can be considered as a lexicographical order [16], where root node contains an empty itemset, and each lower level *k* contains all the *k*-itemsets. Each node of search space is composed of head and tail elements. Head denotes the itemset of node, and items of tail are the possible extensions of new child itemsets. For example with four items *{A, B, C, D}*, in Figure 2 root's head is empty *⟨()⟩* and tail is composed with all of items *⟨(A, B, C, D)⟩*, which generates four possible child nodes *{head ⟨(A)⟩: tail ⟨(BCD)⟩}*, *{head ⟨(B)⟩: tail ⟨(CD)⟩}*, *{head ⟨(C)⟩: tail ⟨(D)⟩}*, *{head ⟨(D)⟩: tail ⟨({})⟩}*. At any node *n*, the candidate N-most *k*-itemset or child nodes of *n* are generated by performing join operation on *n's* head itemset with each item of *n's* tail, and checked for frequency or support counting. This itemset search space can be traversed either by depth first order or breadth first search. At each node, infrequent items from tail are removed by dynamic reordering heuristic [5] by comparing their support with all the itemsets support ($\xi$) and k-itemsets support ($\xi_k$) thresholds [7]. In addition to this, our algorithm also order the tail items by decreasing support which keeps the search space as small as possible [5].

### 3.2. *k*-ITEMSET FREQUENCY CALCUATION

To check, whether any tail item *X* of node *n* at level *k* is N-most interesting *k*-itemset or not, we must check its frequency (support) in *TDS*. Calculating itemset frequency in bit-vector representation requires applying bitwise-$\wedge$ operation on *n.head* and *X* bit-vectors, which can be implemented by a loop; which we call **simple loop**, where each iteration of **simple loop** apply bitwise-$\wedge$ operation on some region of *n.head* with *X* bit-vectors. Since 32-bit CPU supports 32-bit $\wedge$ per operation, hence each region of *X* bit-vector is composed of 32-bits (represents 32 transactions). Therefore calculating frequency of each itemset by using **simple loop** requires applying bitwise-$\wedge$ on all regions of *n*.head with *X* bit-vectors. However, when the dataset is sparse, and each item is present in few transactions, then counting itemset frequency by using **simple loop** and applying bitwise-$\wedge$ on those regions of bit-vectors which contain zero involves many unnecessary counting operations. Since the regions which contain zero, will contribute nothing to the frequency of any itemset, which will be superset of *k*-itemset. Therefore, removing these regions from head bit-vectors (using projection) in earlier stages of search space is more beneficial and useful.

(1) **for** each region index $\ell$ in projected-bit-regions of head
(2) AND-result = bitmap-X [$\ell$] $\wedge$ bitmap-head [$\ell$]
(3) Support-X = Support-X + number of ones(AND-result)

**Figure 3:** Pseudo code of Bit-vector projection using **PBR** technique.

### 3.3. BIT-VECTOR PROJECTION USING PROJECTED-BIT-REGIONS (PBR)

For efficient projection of bit-vectors, the goal of projection should be such as, to bitwise-$\wedge$ only those regions of head bit-vector ⟨*bitmap(head)*⟩ with tail item *X* bit-vector ⟨*bitmap(X)*⟩ that contain a value greater than zero and skip all others. Obviously for doing this, our counting procedure must be so powerful and have some information which guides it, that which regions are important and which ones it can skip. To achieve this goal, we propose a novel bit-vector projection technique **PBR** (Projected-Bit-Regions). With projection using **PBR**, each node *Y* of search space contains an array of valid region indexes $PBR_{(Y)}$ which guides the frequency counting procedure to traverse only those regions which contain an index in array and skip all others. Figure 3 shows the code of itemset frequency calculation using **PBR** technique. In Figure 3, the line 1 is retrieving a valid region index $\ell$ in $\langle bitmap\ (head)\rangle$, while the line 2 is applying a bitwise-$\wedge$ on $\langle bitmap\ (head)\rangle$ with $\langle bitmap\ (X)\rangle$ on region $\ell$.

One main advantage of bit-vector projection using **PBR** is that, it consumes a very small processing cost for its creation, and hence can be easily applied on all nodes of search space. The projection of child nodes at any node *n* can be created either at the time of frequency calculation if pure depth first search is used, or at the time of creating head bit-vector if dynamic reordering is used. The strategy of creating $PBR_{(X)}$ at node *n* for tail item *X* is as; when the **PBR** of $\langle bitmap(n)\rangle$ are bitwise-$\wedge$ with $\langle bitmap(X)\rangle$ a simple check is perform on each bitwise-$\wedge$ result. If the value of result is greater than zero, then an index is allocated in $PBR_{(n.head\ \cup\ X)}$. The set of all indexes which contain a value greater than zero makes the projection of *{n.head $\cup$ X}* node.

### 3.4. MEMORY REQUIREMENT

Some other advantages of projection using **PBR** are that, it is a very scalable approach and consumes very small amount of memory during projection and can be applicable on very large sparse datasets. Scalability is achieved as; we know that by traversing search space in depth first order; a single tree path is explored at any time. Therefore a single **PBR** array for each level of path needs to remain in memory. As a preprocessing step a **PBR** array for each level of maximum path is created and cached in memory. At *k*-itemset generation time different paths of search space (tree) can share this maximum path memory and do not need to create any extra projection memory during itemset mining.

### 3.5. PROJECTION EXAMPLE

Let the second column of Figure 1 (a) shows the transactions of our sample *TDS*. Let the minimum thresholds of $\xi$ and $\xi_k$ for all $(1 \leq k \leq k_{max})$ are equal to 2. For ease of example explanation we will make two assumptions – (a) Each region of bit-vectors contain only a single bit (b) A static alphabetical order will be followed instead of ascending frequency order.

At the start of algorithm, the bit-vector of each single item in the dataset is created. Then, the N-most interesting itemset mining process from the root node is started. At root node, head itemset is *{}*, therefore the PBR of each tail item $X = (A, B, C, D, E, I)$ is created by traversing all $(bitmap(X))$ regions, and indexing only those which contain a value greater than 0. For example $(bitmap\ (A))$ contains a value greater than zero in bit regions $((01, 02, 05, 06, 07))$; thereby these are its PBR *indexes*. Since we are traversing search space in depth first order and item *{A}* is first in alphabetical order. Therefore, root node creates a child node with *head* $((A))$ and *tail* $((B, C, D, E, I))$ and iterates to *node {A}*. At *node {A}*, frequency of each tail item *X* is calculated by applying bitwise-∧ on $(bitmap\ (A))$ with $(bitmap(X))$ on $PBR_{(A)}$. Figure 4 (a) shows the frequency calculation process of itemset $((X \cup head.\{A\}))$ by using PBR. After frequency calculation tail items $((B: 4, C: 2, D: 2))$ are found locally frequent at *node {A}*. Since item *{B}* in *tail.{A}* is next in alphabetical order, therefore the *node {A}* creates a new child node with *head* $((AB))$ and *tail* $((C, D))$. PBR = $((01, 02, 05, 06))$ of child *node {AB}* at *node {A}* are created by applying bitwise-∧ on $((bitmap\ (A)))$ with $((bitmap\ (B)))$ on $PBR_{(A)}$. The mining process then moves to *node {AB}* with *head* $((AB))$ and *tail* $((C, D))$.

Similarly at *node {AB}*, frequency of each tail item *X* is calculated by applying bitwise-∧ on $(bitmap\ (AB))$ with $(bitmap(X))$ on $PBR_{(AB)}$. Figure 4 (b) is showing the frequency calculation process at *node {AB}*. After frequency calculation tail items $((C: 2, D: 2))$ are found locally frequent at *node {AB}*. Since item *{C}* in *tail.{AB}* is next in alphabetical order, therefore the mining process creates a child node with *head* $((ABC))$ and *tail* $((D))$. PBR = $((01, 06))$ of child *node {ABC}* at *node {AB}* are created by applying bitwise-∧ on $(bitmap\ (AB))$ with $((bitmap\ (C)))$ on $PBR_{(AB)}$. Afterward, the mining process iterates to *node {ABC}* with *head* $((ABC))$ and *tail* $((D))$.

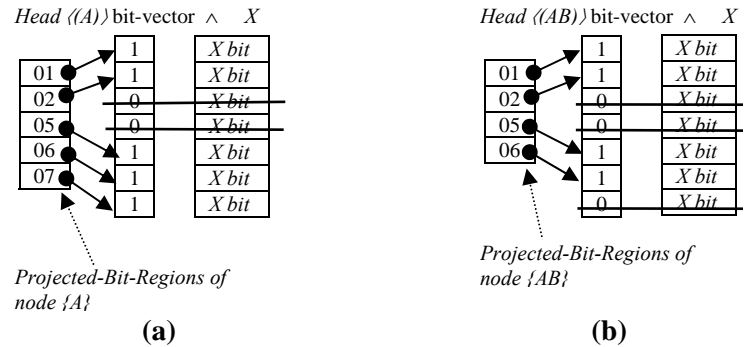

**Figure 4:** (a) Itemset *(AB ∪X)* frequency calculation using **PBR.** (b) Itemset *(AB ∪X)* frequency calculation using **PBR**.

At *node {ABC}*, all tail items are infrequent; therefore the mining process stops and backtracks at *node {AB}* and mine other items of its *tail ((D))*. Since *item {D}* is next in alphabetical order, so the mining process creates a new child *node {ABD}* with *head ((ABD))* and *tail (({}))*. At *node {ABD}*, tail is empty so the mining process stops and backtracks at *node {A}*. Similarly, the mining process mines other items of *node {A}* tail in same fashion. After mining all tail items of *node {A}*, the mining process then backtracks at root node, and mines other tail items of root in depth first order. Due to the lack of space we could not give a detailed example, but the basic idea for mining other itemsets is similar.

### 3.6. ADJUSTING THRESHOLD $\xi_k$ AND $\xi$ IN N-Most FREQUENT ITEMSET MINING

At the start of algorithm, we set the minimum support threshold of all the *k*-itemsets $\xi_k$ *($1 \leq k \leq k_{max}$)* to zero. With $\xi_k = 0$ we would blindly include any *k*-itemsets in our current set of N-most k-itemset. Once the algorithm encountered any Nth *k*-itemset for some level $k \leq k_{max}$, we can safely set the minimum support threshold of $\xi_k$ by assigning to it the minimum value among the support of the N-most *k*-itemset discovered so far. Next, only that *k*-itemset will be added in the N-most *k*-itemsets result, which will contain support greater than $\xi_k$. Once the algorithm finds such *k*-itemset, the support threshold of $\xi_k$ is adjusted again by assigning to it the minimum value among the support of the N-most *k*-itemsets.

In the mining phase, we set the initial threshold value of $\xi$ (all the itemsets support) to zero. During mining, the algorithm increases $\xi$ by assigning to it the minimum value among the supports of the Nth most frequent *k*-itemset discovered so far for $\xi = min\ (\xi_1,\ \xi_2, ....\ \xi_{kmax})$. Figure 5 shows the pseudo code of N-MostMiner.

```
N-MostMiner (Node n)

(1)  for each item X in n.tail
(2)     for each region index ℓ in PBR₍n₎
(3)        AND-result = bit-vector[ℓ] ∧ head_bit_vector of n [ℓ]
(4)        Support[X] = Support[X] + number_of_ones(AND-result)

(5) Remove infrequent items from n.tail, if support less than ξₖ or ξ
(6) Reorder them by decreasing support

(7) for each item X in n.tail
(8)     m.head = n.head ∪ X
(9)     m.tail = n.tail – X
(10)    for each region index ℓ in PBR₍n₎
(11)    AND-result = bit-vector[ℓ] ∧ head-bit-vector[ℓ]
(12)        If AND-result > 0
(13)            Insert ℓ in PBR₍m₎
(14)            head bit-vector of m [ℓ] = AND- result
(15)    N_most_k_itemset = N_most_k_itemset ∪ n.head
(16)    if N_most_k_itemset == Nth itemset
(17)        increase the support ξₖ by assigning minimum value support of N_most_k_itemset
(18)    N_MostMiner (m)
```

**Figure 5:** Pseudo code of mining N-Most all frequent itemsets

### 3.7. MINING TOP-K FRQUENT CLOSED ITEMSET

In mining Top-K frequent closed itemsets, we incorporate the same methodology for itemset generation, frequency counting and projection, as we have described for N-Most all frequent itemset mining in above sections. The only one big difference is that, a frequent closed itemset is considered to be Top-K only when none of its superset found to be Top-K frequent closed itemset. This closed frequent itemset superset checking can be implementation using a straightforward sequential searching method, and could take large processing time if the K size is given to be very large. In mining Top-K frequent closed itemset there is no concept of k-itemset support threshold at each level, therefore only the minimum support threshold of all the itemset $\xi$ is set to zero. With $\xi = 0$ we would blindly include any itemsets in our current set of Top-k frequent itemset with minimum length greater than min_l. Once the algorithm encountered any k-th itemset, we can safely set the minimum support threshold of $\xi$ by assigning to it the minimum value among the support of the Top-k frequent itemsets discovered so far. Next only that itemset will be added in the Top-k frequent itemsets result, which will contain the support and minimum length greater than $\xi$ and min_l. Figure 6 shows the pseudo-code of mining Top-K-Miner.

---
*Top-k-ClosedMiner* (Node n)
---

*(1) for each item X in n.tail*
*(2)   for each region index ℓ in $PBR_{\langle n \rangle}$*
*(3)     AND-result = bit-vector[ℓ] ∧ head_bit_vector of n [ℓ]*
*(4)     Support[X] = Support[X] + number_of_ones(AND-result)*

*(5) Remove infrequent items from n.tail, if support less than ξ*
*(6) Reorder them by decreasing support*

*(7) for each item X in n.tail*
*(8)   m.head = n.head ∪ X*
*(9)   m.tail = n.tail – X*
*(10)  for each region index ℓ in $PBR_{\langle n \rangle}$*
*(11)  AND-result = bit-vector[ℓ] ∧ head-bit-vector[ℓ]*
*(12)     If AND-result > 0*
*(13)       Insert ℓ in $PBR_{\langle m \rangle}$*
*(14)       head bit-vector of m [ℓ] = AND- result*
*(15)  Top-k-ClosedMiner (m)*
*(16)  if none of (n.head) superset itemset is frequent closed Top-K itemset*
*(18)     Top_k_List = Top_k_List ∪ n.head*
*(19)     if Top_k_List == kth itemset*
*(20)       increase the support ξ by assigning minimum value  support of Top_k_List*

**Figure 6:** Pseudo code of mining Top-K frequent closed itemsets

## 4. EFFICIENT IMPLENATION TECHNIQUES

In this section, we provide some efficient implementations ideas, which we experienced in our N-MostMiner and Top-K-Miner implementations.

### 4.1. ELIMINATION REDUNDANT FREQUENT COUNTING OPERATIONS (ERFCO)

The codes which we describe in Figure 5 and Figure 6 performs exactly two frequency counting operations for each frequent tail item *X* at any node *n* of search space. First, at the time of performing dynamic reordering, and second, for creating *{X∪ n.head}* bit-vector. The itemset frequency calculation process which is considered to be the most expensive task (penalty) in overall itemset mining [10], the bit-vector representation approach suffers this penalty twice for each frequent *k*-itemset. The second counting operation which we can say is redundant, occurs due to gain efficiency in 32-bit CPU and can be eliminated with some efficient implementation, which we describe below.

In N-MostMiner, at the start of algorithm two large heaps, one for head bit-vectors and one for PBR are created (with 32-bit per heap slot size). Next, at the time of calculating frequency of *k*-itemset *X* a simple check is performed to ensure that is there sufficient space left in both heaps. If the response is "yes" then the head bit-vector of *X* and *PBR*$_{(X)}$ are created at the same time when dynamic reordering is performed, otherwise normal procedure is followed. The main difference is that, with the efficient implementation bitwise-∧ results and regions indexes are written in heaps instead of tree path levels memories. The size of heaps should be so enough that it can store any frequent item subtree. From our implementation point of view, we suggest that heap size double the total number of transactions is enough for very large sparse datasets. In our N-MostMiner and Top-K-Miner implementations we found that it completely eliminates the second frequency counting operation while requiring very little amount of memory.

### 4.2. INCREASING PROJECTED BIT-REGIONS DENISTY (IPBRD)

The bit-vector projection technique which we described in section 3.3 does not provide any compaction or compression mechanism for increasing the density in items bit-vector regions. As a result, on the sparse dataset only one or two bits in each region of item bit-vector are set to one, which not only increases the projection length but also with this, it is not possible to achieve true 32-bit CPU performance. Therefore, for increasing the density in bit-vector regions the N-MostMiner and Top-K-Miner starts with an array-list [15]. Next at root node, a bit-vector representation for each frequent item is created which provides sufficient compression and compaction in bit-vectors regions. Sufficient improvements are obtained in our algorithms by using this approach.

### 4.3. 2-ITEMSET PAIR

There are two methods to check whether current candidate *k*-itemset is frequent or infrequent. First, to directly compute its frequency from *TDS*. Second one, which is more efficient, is known as 2-Itemset pair. If any 2-Itemset pair of any candidate *k*-itemset is found infrequent, then by following Apriori property [1], the candidate *k*-itemset will be also infrequent. We know any *k*-itemset which contains a length more than two, is the superset of its entire 2-Itemset pairs. Therefore, before counting its frequency from transactional dataset, our algorithms check its 2-Itemset pairs. If any pair is found infrequent (support less than $\xi_k$ *or* $\xi$), then that *k*-itemset is automatically considered to be infrequent without checking its frequency in *TDS*.

# 5. PERFORMANCE EVALUATION

In this section we report our performance results of N-MostMiner and Top-K-Miner on a number of benchmark datasets. For experimental purpose we used the original source code of BOMO, which is freely available at http://www.cse.cuhk.edu.hk/~kdd/program.html. Unfortunately, there is no publicly available implementation of TFP, so in experiments we used our own implementation which is written in C language. All the source codes of N-MostMiner and Top-K-Miner are written in C language. The experiments are performed on 3.2 GHz processor with main memory of size 512 MB, running windows XP 2005 professional. For experiments, we used the benchmark datasets available at http://fimi.hi.cs.datasets. These datasets are frequently used in many frequent itemset mining algorithms. For detailed performance evaluation we classified the datasets into four different groups and select one or two datasets from each group. Our first group is composed of BMS-WebView1, BMS-WebView2 and Retail datasets. These datasets have a large number of items but small number of transactions and are sparse. We choose BMS-WebView2 for performance comparison. Our second group is composed of BMS-POS and Kosarak datasets. These datasets have many items and also large number of transactions. If the minimum support is given to be very small, these datasets generates huge number of frequent itemsets. We chose BMS-POS for performance comparison. Our third group is composed of Chess, Connect, Pumsb, Pumsb-star, accidents and Mushroom datasets. These datasets are very dense and almost 90% of time is spend on writing frequent itemsets to output file, if the minimum support is given to be very small. We choose Mushroom and Chess for performance comparison. Our last group is composed of T10I4D100K and T40I10D100K. These datasets are very sparse and have large number of items. We select both datasets for performance comparison. Table 1 shows the description of datasets that we used in our experiments.

| Dataset | Items | Average Transactional Length | Records |
|---|---|---|---|
| T10I4D100K | 1000 | 10 | 100,000 |
| T40I10D100K | 1000 | 40 | 100,000 |
| Chess | 75 | 35 | 3,196 |
| Mushroom | 119 | 23 | 8,124 |
| BMS-POS | 1658 | 7.5 | 515,597 |
| BMS-WebView2 | 3341 | 5.6 | 77,512 |

**Table 1:** Datasets used in the our computational experiments

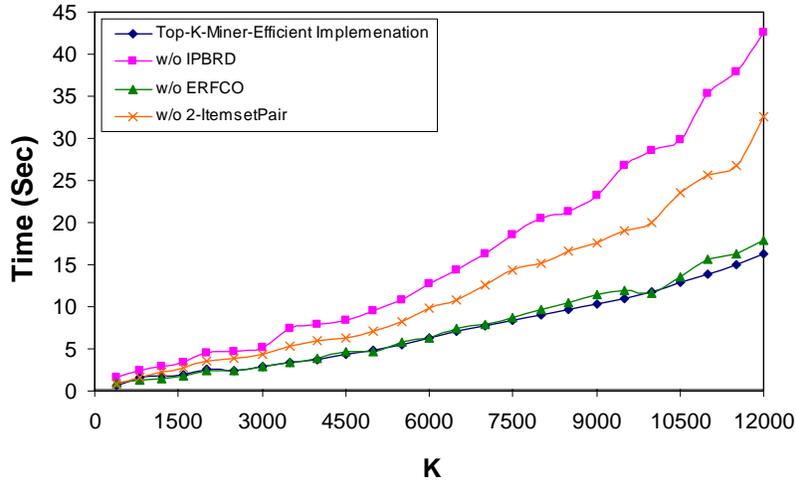

**Figure 7:** Performance results of efficient implementation components on T1014D100K dataset

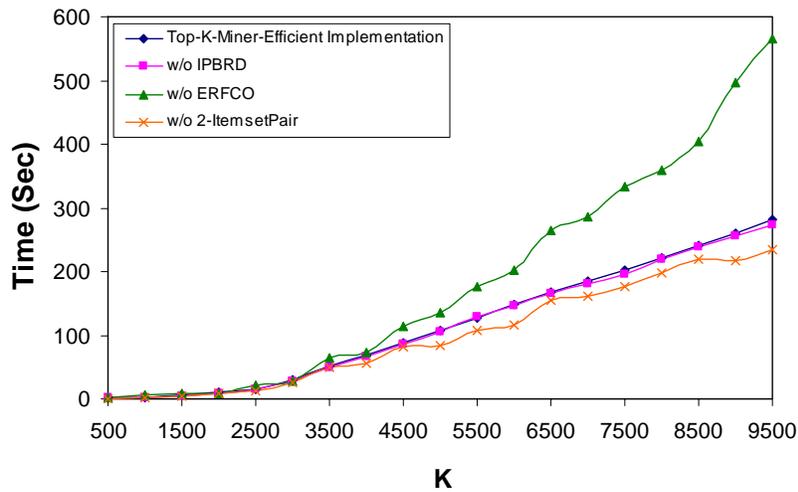

**Figure 8:** Performance results of efficient implementation components on chess dataset

## 5.1. PERFORMANCE ANALYSIS OF EFFICIENT IMPLEMENTATION COMPONENTS

In this section, we show the performance analysis and effect of each component of our efficient implementation techniques on sparse and dense type datasets. The simulations are performed only on mining Top-K frequent closed itemset problem, but we are sure that each component also has exactly same effect on mining N-Most all frequent itemset problem. Our first experiment is on T10I4D100K, which is sparse dataset. Figure 7 shows the experimental results on T10I4D100K dataset. From the Figure result, it is clear that IPBRD has the biggest effect on low level support thresholds as compared to other three components. On higher level threshold, we paid some extra cost of 2-Itemset pair checking, but as the support threshold decreases, 2-Itemset pair checking improves the overall performance of algorithm. Our second experiment is on Mushroom, which is a dense dataset. Figure 8 shows the experimental results on Mushroom dataset. On dense dataset it is cleared that IPBRD and 2-Itemset pair does not create any major effect, due to small number of items and transactions. On dense datasets, only the ERFCO and itemset frequency counting using Bit-vector dominate the overall algorithm execution performance.

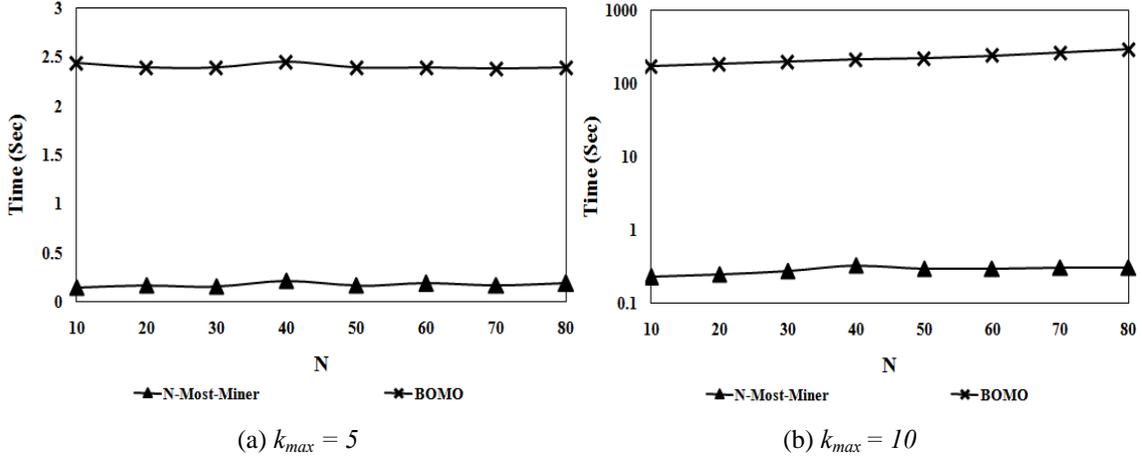

(a) $k_{max} = 5$      (b) $k_{max} = 10$

**Figure 9:** Performance results on Chess dataset

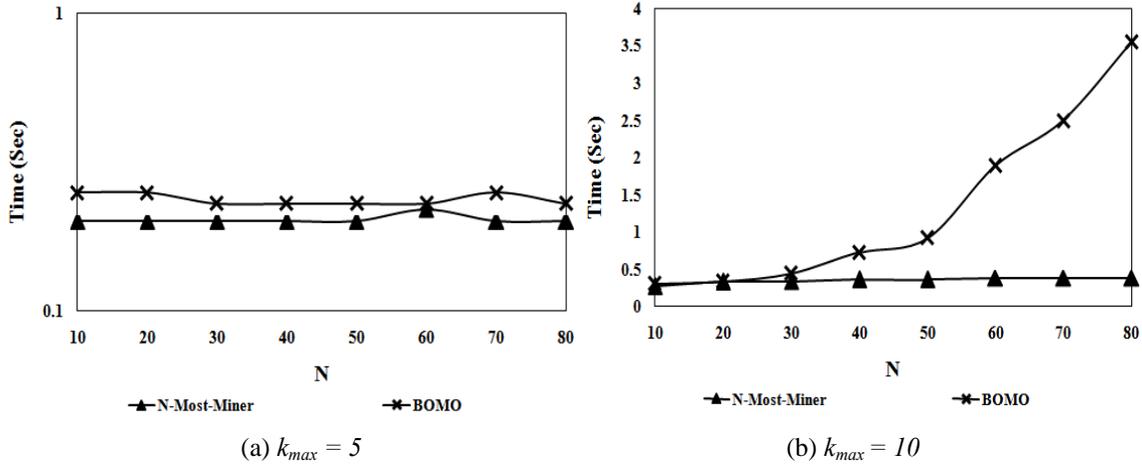

(a) $k_{max} = 5$      (b) $k_{max} = 10$

**Figure 10:** Performance results on Mushroom dataset

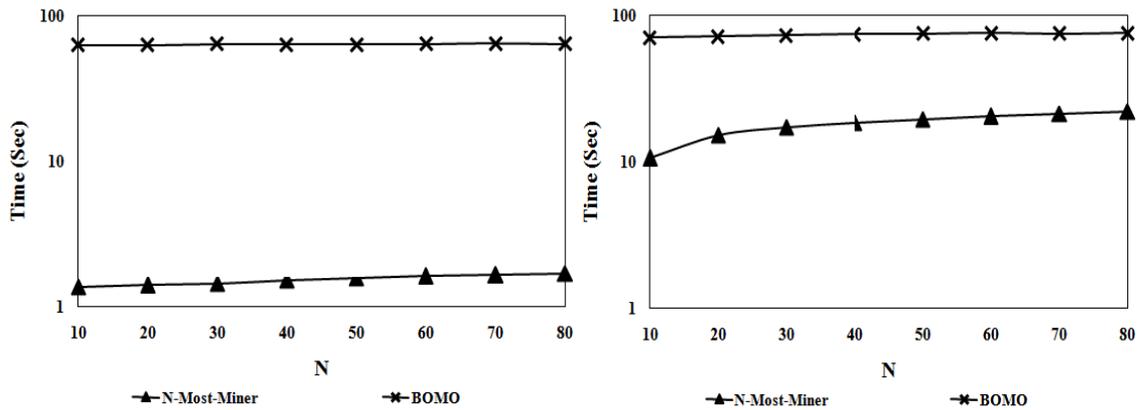

(a) $k_{max} = 5$  (b) $k_{max} = 10$

**Figure 11:** Performance results on T1014D100K dataset

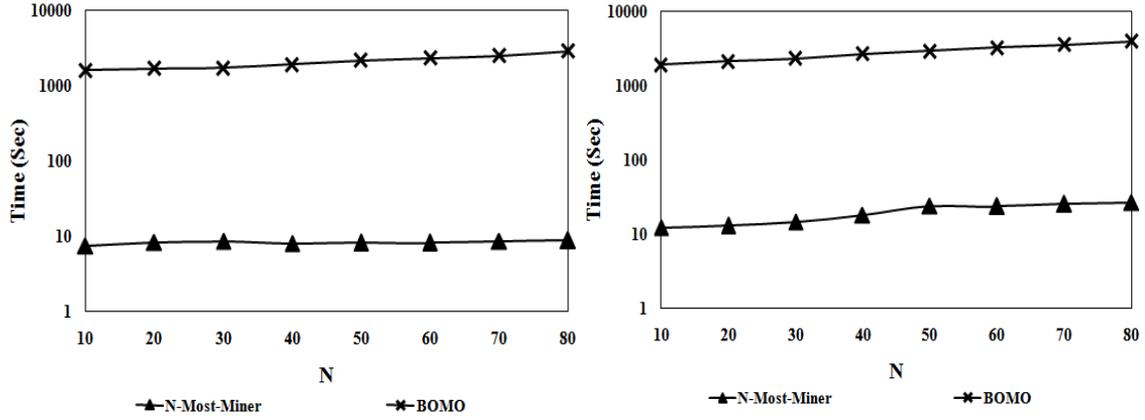

(a) $k_{max} = 5$  (b) $k_{max} = 10$

**Figure 12:** Performance results on T40I10D100K dataset

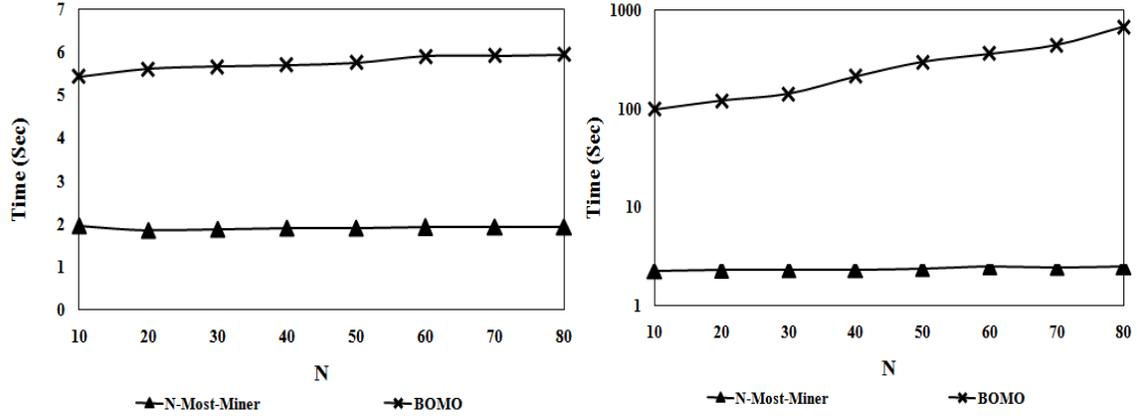

(a) $k_{max} = 5$  (b) $k_{max} = 10$

**Figure 13:** Performance results on BMS-WebView2 dataset

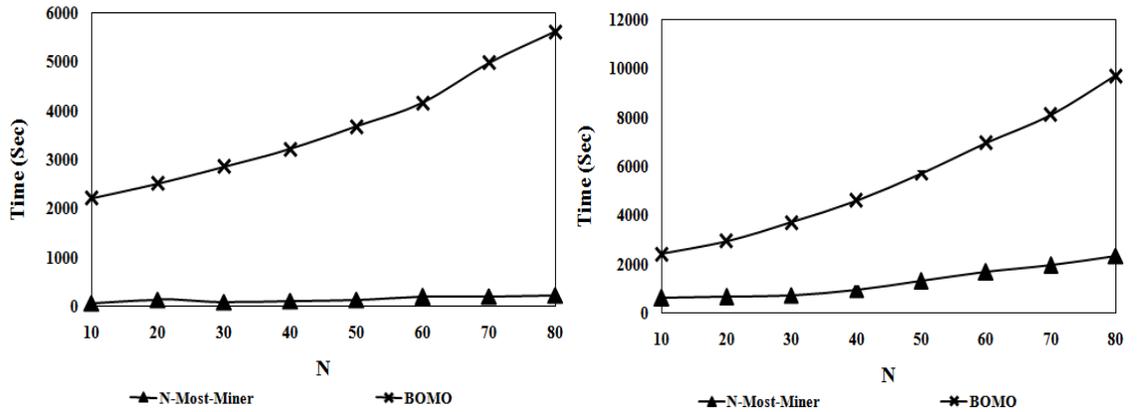

(a) $k_{max} = 5$  (b) $k_{max} = 10$

**Figure 14:** Performance results on BMS-POS dataset

## 5.2. PERFORMANCE EVALUATION OF N-Most-Miner

We perform our computational experiments using the different N-most values 10, 20, 30, 40, 50, 60, 70 and 80 under two $k_{max}$ thresholds values, $k_{max} = 5$ $((1 \leq k \leq 5)$ and $k_{max} = 10$ $((1 \leq k \leq 10)$. The performance measure is the execution time of the algorithms under different N and $k_{max}$ threshold values. Figures 9-14 results show that the N-MostMiner outperforms the BOMO on both all and closed frequent itemsets mining problems, on almost all levels of mining thresholds on all types of sparse and dense datasets. This is due to its fast frequency counting of $k$-itemset, projection and efficient implementation techniques. Whereas the BOMO requires several indirect memory accesses during each $k$-itemset frequency counting, which slows down the whole mining process.

Figure 9 and 10 show the performance results of two algorithms on dense type datasets (chess and mushroom). These datasets contain a few numbers of items but a maximal transactional length which is same in almost all transactions. Due to the small number of transactions, the performance result of $k$-itemset frequency counting using bit-vector representation approach is more efficient as compared to pattern-growth approach. Although in these datasets a large number of transactions share a common path in FP-tree. Figure 11 to 14 show the performance results of two algorithms on four sparse type datasets both real and synthesized. Again the N-MostMiner outperforms the BOMO algorithms on almost all type of mining thresholds on all type of sparse datasets. On experiments with sparse type datasets we note that BOMO takes a much larger time for its initial FP-tree construction, when $\xi_k$ $(1 \leq k \leq k_{max})$ and $\xi$ are both equal to zero. In fact, on sparse type datasets N-MostMiner (in most of the cases) finishes its execution, when BOMO busy in its initial FP-tree construction. The reason is that, on sparse datasets with large number of items and small average transaction length, most of the transactions can't share a common prefix path in FP-Tree that results in a large initial FP-Tree construction, which slows down the itemset frequency counting. The performance results of our efficient implementation techniques of N-MostMiner especially 2-Itemset Pair and IPBRD are more encouraging on sparse type datasets as compared to dense type datasets.

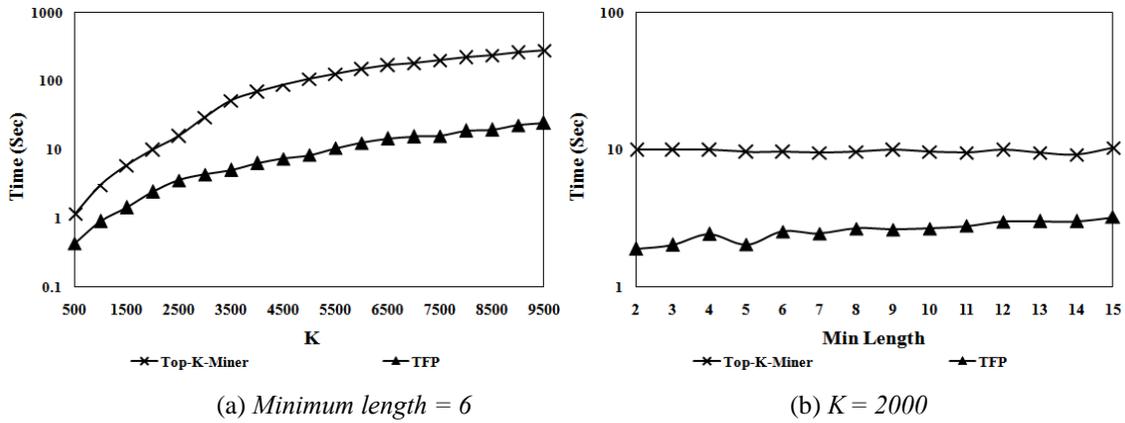

(a) *Minimum length = 6*   (b) *K = 2000*

**Figure 15:** Performance results of Top-K-Miner on Chess dataset

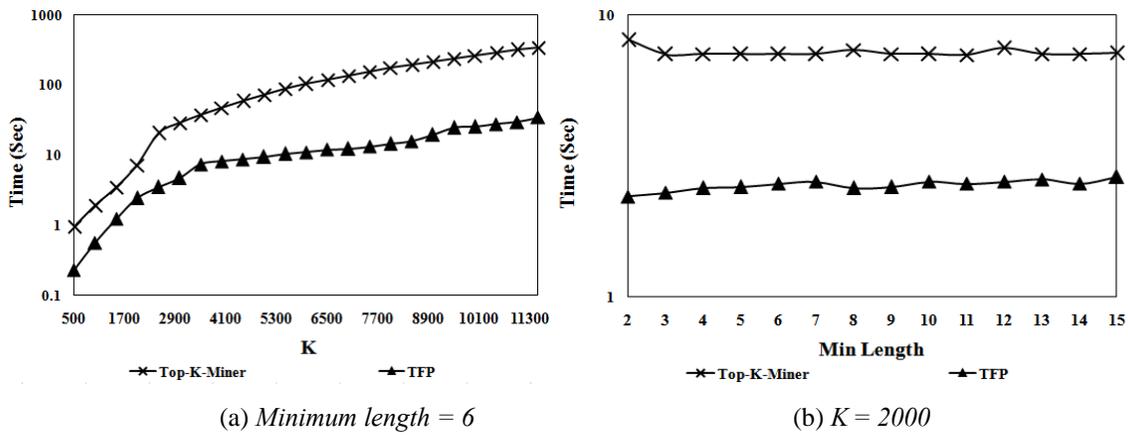

(a) *Minimum length = 6*   (b) *K = 2000*

**Figure 16:** Performance results of Top-K-Miner on Mushroom dataset

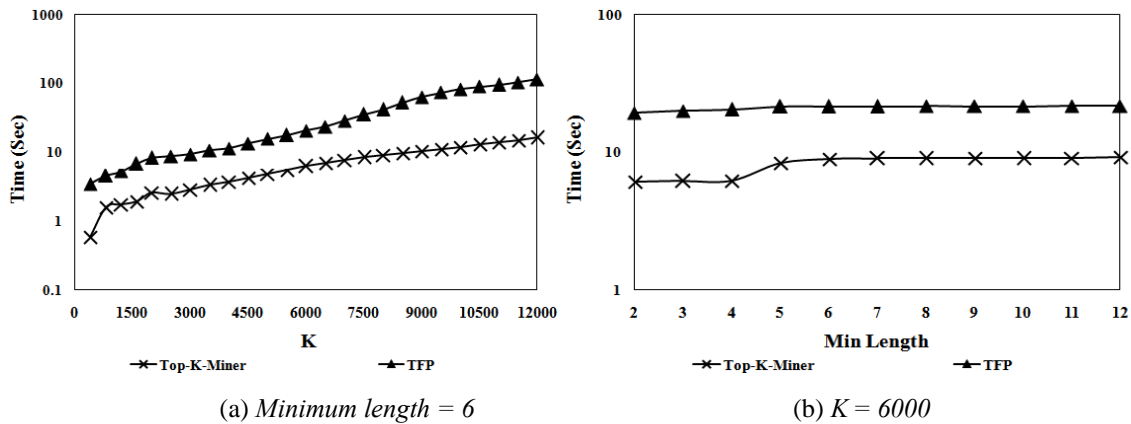

(a) *Minimum length = 6*   (b) *K = 6000*

**Figure 17:** Performance results of Top-K-Miner on T1014D100K dataset

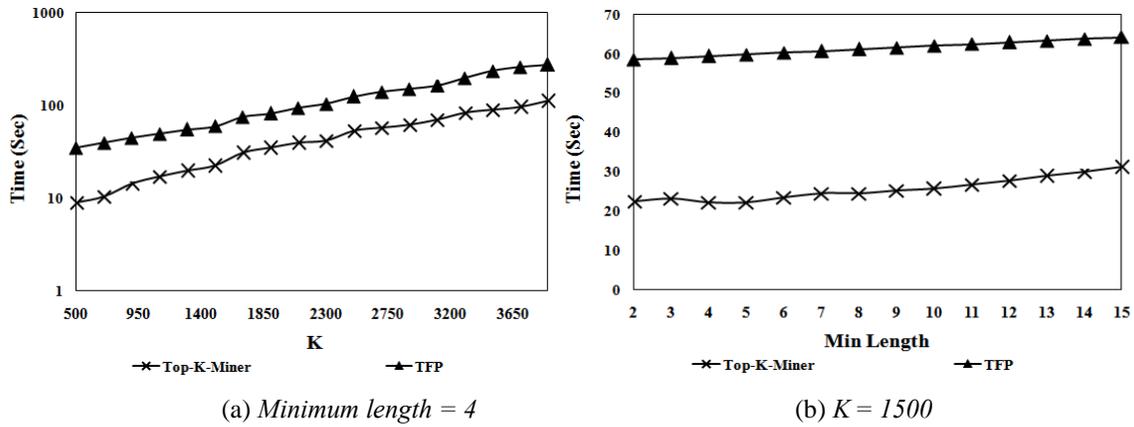

(a) *Minimum length = 4*        (b) *K = 1500*

**Figure 18:** Performance results of Top-K-Miner on T40I10D100K dataset

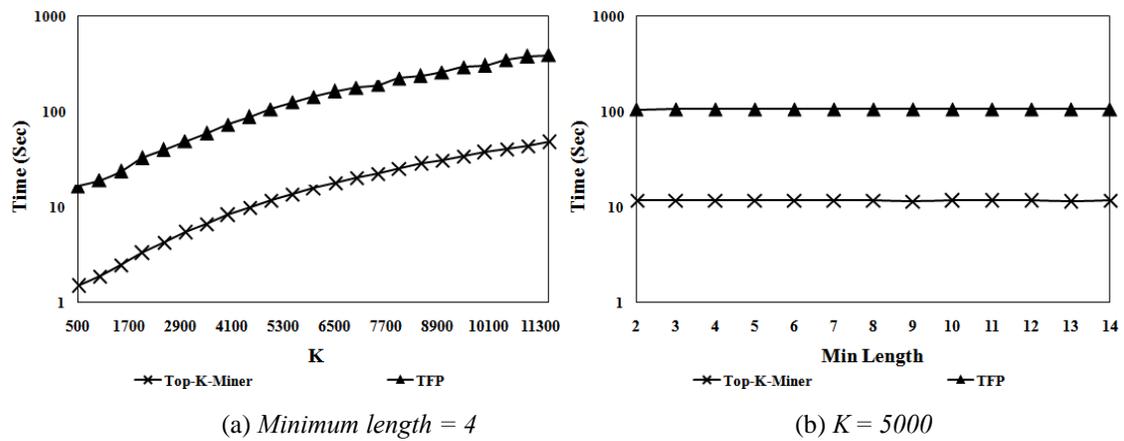

(a) *Minimum length = 4*        (b) *K = 5000*

**Figure 19:** Performance results of Top-K-Miner on BMS-WebView2 dataset

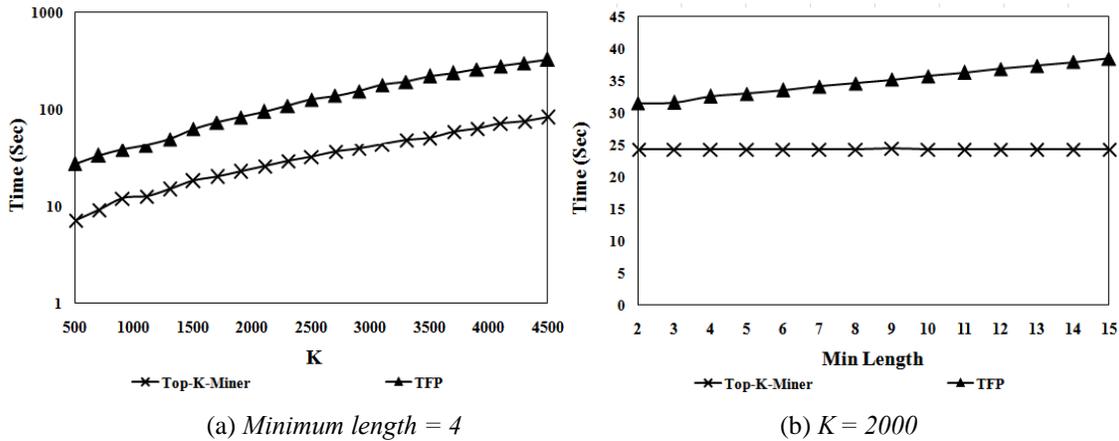

(a) *Minimum length = 4*        (b) *K = 2000*

**Figure 20:** Performance results of Top-K-Miner on BMS-POS dataset

## 5.3. PERFORMANCE EVLATUATION OF TOP-K-MINER

Our second set of experiments were on mining Top-K closed frequent itemsets mining with minimum length greater than *min_l* threshold. The experiments were performed on each dataset using two different scenarios. In first, we fixed the minimum length and varied the K value. While in second, the K value was kept fixed and minimum length was varied. Figure 15 and 16 show the performance result of TFP and Top-K-Miner on two dense type datasets (Chess and Mushroom). Due to small number of items and large average transactional length, TFP creates a compact initial FP-Tree, in which lot of transactions shares a common prefix path, which helps in fast itemset frequency counting. On dense type datasets, we note that the efficient implementation techniques of Top-K-Miner especially 2-Itemset Pair and IPBRD does not create any major performance effect, only frequency counting dominates the whole algorithm execution result.

Figure 17 to 20 show the performance results on two algorithms on sparse type datasets. As clear from the figure results the Top-K-Miner outperform the TFP on almost all levels of mining threshold values, due to its efficient bit-vector projection and implementation techniques especially 2-Itemset Pair and IPBRD. On spare datasets with large number of items and transactions, TFP faces the same problem as BOMO of its initial FP-Tree construction, when $\xi$ equal to zero, which slows down the whole algorithm execution.

## 6. CONCLUSTION

Mining interesting frequent itemsets without minimum support threshold are more costly in terms of processing time due to large area of itemset search space exploration as compared to traditional frequent itemset approach. Due to large number of candidate itemsets generation, the efficiency of mining interesting (N-Most or Top-K) frequent itemsets algorithm highly depends upon the two main factors – (a) Dataset representation approach for fast frequency counting – (b) Projection of relevant transactions to lower level nodes of search space. In this paper we present two novel algorithms for mining interesting frequent itemset (N-MostMiner and Top-K-Miner) using bit-vector representation approach, which is very efficient in terms of candidate itemset frequency counting. For projection we present a novel bit-vector projection technique PBR (projected-bit-regions), which is very efficient in terms of processing time and memory requirement. Several efficient implementation techniques of N-MostMiner and Top-K-Miner are also presented, which we experienced in our implementation. Our experimental results on benchmark datasets suggest that mining interesting frequent itemsets without minimum support threshold using N-MostMiner or Top-K-Miner is highly efficient in terms of processing time as compared to currently best algorithms BOMO and TFP. This shows the effectiveness of our algorithm.